\documentclass[preprint,12pt]{elsarticle}

\usepackage{amssymb}
\usepackage{amsmath}
\usepackage{ulem}

\usepackage{xcolor}

\journal{Journal of Subatomic Particles and Cosmology}

\begin{document}
\begin{frontmatter}

\title{Energy density in color string clusters and its implications for heavy-quark energy loss}

\author[a]{Daria Prokhorova}
\author[a]{Shuzhe Shi}
\author[b]{Evgeny Andronov}
\affiliation[a]{organization={Tsinghua University},
             addressline={Haidian  District},
             city={Beijing},
             postcode={100084},
             country={China}}

\affiliation[b]{organization={St. Petersburg State University},
             addressline={Peterhof},
             city={St. Petersburg},
             postcode={198504},
             country={Russia}}

\begin{abstract}
Initial stages of relativistic ion and proton-proton collisions can be described within the framework of color flux tube formation between the receding color charges. In the approximation of fully penetrable color string volumes, regions of high string density yield overlaps of randomly oriented color fields and create string clusters. The paper discusses two ways of characterizing the medium formed by the resulting field configurations. One is based on the chromo-electric field strength that scales as the square root of the number of color sources in the cluster. The other is defined by the energy density of the confined color field and thus scales linearly with that number. We compare the effect of the two approaches on the extracted effective temperature of the medium, which is the input for calculations of collisional and radiative energy loss of a probe traversing it.
\end{abstract}

\end{frontmatter}


\section{Introduction}

The early stages of relativistic ion-ion and proton-proton collisions are characterized by extreme conditions. At soft momentum scales and high parton densities, non-perturbative models are required to describe the initial state that defines the subsequent evolution of the created matter.

In the Color Glass Condensate framework~\cite{Gelis:2010nm}, the high density of gluons at small $x$ leads to large occupation numbers, so the gluon field is treated as a classical chromo-electric and chromo-magnetic field. Its transverse correlation length is set by the saturation scale $Q_s$ and naturally determines the characteristic transverse size $r_\perp\sim1/Q_s$ of the glasma fields~\cite{Gelis:2012ri} formed immediately after a collision. When the same dense field regions are described in terms of string-like degrees of freedom, the color rope model~\cite{Biro:1984cf} treats overlapping sources as forming a $K$-fold rope. Assuming the rope thickness is independent of its foldness, the field strength of the rope scales as $\sqrt{K}$ due to the random walk in $SU(3)$ color space. In the Color String Percolation Model (CSPM)~\cite{Braun:2015eoa}, strings are represented by fully penetrable disks of finite transverse area, whose overall density is characterized by a percolation parameter~\cite{Braun:1997ch}, while their overlap regions form local clusters. Here, the $\sqrt{k}$ dependence emerges for the cluster charge due to averaging over the random orientations of $k$ color fields. This scaling governs both the average multiplicity per unit rapidity and the average squared transverse momentum of particles produced by Schwinger-like decay of such a $k$-string cluster~\cite{Braun:1999hv}.

Alternatively, the color field configuration can be converted into a medium by matching its energy-momentum tensor to that of a plasma~\cite{Fries:2005yc}. In our previous work~\cite{Prokhorova:2025jmd}, we relaxed the assumption of the full field-to-plasma correspondence of $T_{\mu\nu}$ and performed only the thermal matching of the energy density $T_{00}$. Assuming that the energy density of a $k$-string cluster available for medium formation inherits the $\sqrt{k}$ scaling of the color field strength, we extracted the effective local temperature of an equilibrated gluon medium with the same local energy density.

In this work, we instead identify the energy density of the $k$-string cluster directly with the field energy, which for the same setup of randomly oriented color charges scales as $k$ rather than $\sqrt{k}$. We contrast this thermodynamic energy density with the field-strength scaling that controls particle production, and estimate the resulting difference in the extracted effective temperature. We also evaluate the average fraction of a heavy-flavor probe's trajectory that lies in dense string regions and compute the jet-path-averaged temperature along it. The latter sets the magnitude of the medium-induced energy loss, which will be presented in detail elsewhere. Section~\ref{sec2} describes our analysis, and Section~\ref{sec3} discusses the results.

\section{Medium energy density from string overlaps}
\label{sec2}
In the color string picture~\cite{Artru:1974hr}, the number of color flux tubes formed between receding color charges can be estimated from a Glauber model, a dipole picture, or a parametrized distribution of multi-pomeron exchanges~\cite{Capella:1992yb,Kaidalov:1982xe}. The latter produces a substantial number of strings even in $p+p$ collisions at high energies.

To keep the numerical simulations computationally feasible, strings are typically approximated as having a sharp boundary, represented either as circles on the transverse plane (infinite rapidity strings) or as 3D cylinders (finite-length strings). As a result, string intersections yield local clusters of varying overlap order $k$. Within this effective picture, the total charge of a $k$-string cluster is given by the color sum rule, $Q_k = \sqrt{k}Q_1S_k/S_1$, where $S_k$ is the area of the overlap region, while $S_1$ and $Q_1$ are the area and the charge of a single string, respectively~\cite{Braun:1991dg}. The field strength is then given by $E_k = Q_k/S_k$~\cite{Lyakhov:2011dzl}, so that $E_k = \sqrt{k}E_1$. This resulting field strength enters the Schwinger pair-production exponent~\cite{Schwinger:1951nm}, $\exp \left[ -\pi(m^2 + p_\perp^2)/{|eE_k|} \right]$. Following the phenomenological string fusion prescription, the mean particle multiplicity per unit rapidity is assumed to be proportional to the effective color charge of the emitting region, yielding, for a color source of $k$ overlap order, $\mu_k= \mu_1\;\sqrt{k}\;S_k/S_1$, where $\mu_1$ is that from a single string~\cite{Garcia:2023zfx}.

In our previous work~\cite{Prokhorova:2025jmd}, we assumed that the energy density effectively inherits the field-strength scaling, $\varepsilon_k \propto E_k \propto \sqrt{k}$. However, from the thermodynamic point of view, the energy density of the chromo-electric field is $\varepsilon_k = \langle \vec{E}^2_k \rangle/2 = \sum^k_{i,j=1}\langle \Vec{E}_i
\cdot \Vec{E}_j \rangle /2$. In our simulation, we do not track the explicit color state of each string. Instead, we treat the color orientations as random and average over them analytically. Thus, the cross terms ($i \neq j$) average to zero, simplifying the expression to $\varepsilon_k = E_k^2/2$ and yielding the local energy density $\varepsilon_k = k\,\varepsilon_1$ for the $k$-order overlap region, where $\varepsilon_1 = \sigma/S_1$ is the energy density of a single string, $\sigma = 0.9$~GeV/fm~$\approx 0.18$~GeV$^2$ is the string tension, and the single-string area is $S_1 = \pi r_0^2$ with radius $r_0 = 0.25$~fm. In both $\varepsilon \propto \sqrt{k}$ and $\varepsilon \propto k$ cases, the effective local temperature of an ideal gluon gas with the same energy density as that of the $k$-string cluster is given by $T_k = \left(30\;\varepsilon_k/16\pi^2\right)^{1/4}$. We neglect the contribution of light quarks, following~\cite{Prokhorova:2025jmd}, which would increase the degeneracy factor $g=16$ and lower the extracted temperature uniformly in both scalings. 

We simulate $p+p$ collisions at $\sqrt{s} = 5.02$ TeV with the number of infinite-length strings $N_s$ ranging from 2 to 100, which corresponds to roughly $7$ orders of magnitude in event probability~\cite{Prokhorova:2025jmd}. In each event, the centers of $N_s$ disks are sampled from a 2D Gaussian of width $0.6$~fm inside the $4 \times 4$~fm$^2$ area. The string configuration is assumed to be static. We also neglect Bjorken expansion and the corresponding decrease in energy density.

\begin{figure}[h!]
\centering
\includegraphics[width=1.0\linewidth]{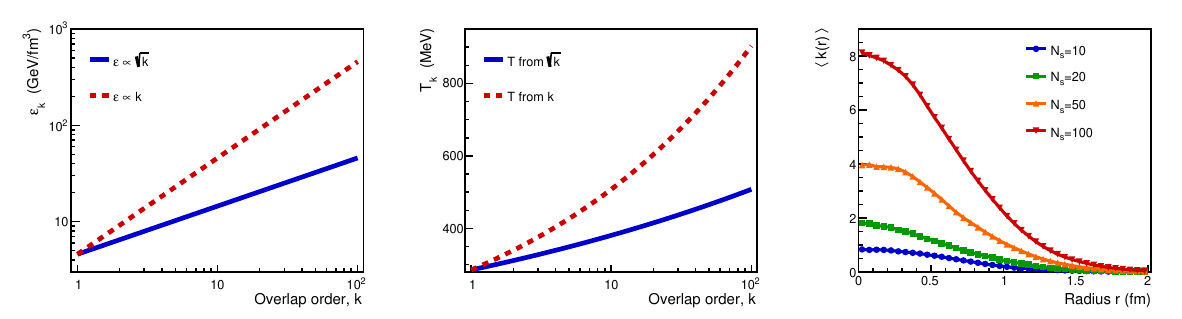}
\caption{Energy density (left) and effective temperature (middle) versus overlap order $k$ for the two energy-density scalings. Right: average overlap order $\langle k(r) \rangle$ versus radial distance from the origin for different total number of strings $N_s$ in event.}\label{fig1}
\end{figure}

\begin{figure}[h!]
\centering
\includegraphics[width=1.0\linewidth]{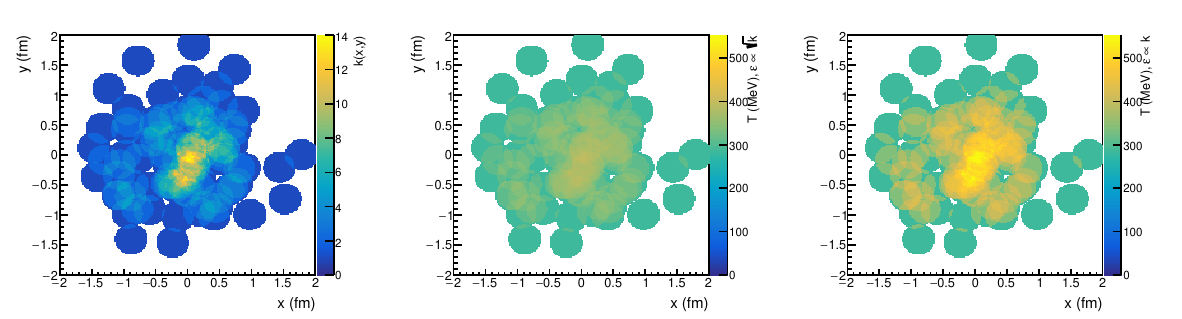}
\caption{Transverse slices of a rare $p+p$ event with $N_s = 100$ strings. Shown are the number of overlaps per cell (left) and the resulting temperatures for the $\varepsilon \propto \sqrt{k}$ (middle) and $\varepsilon \propto k$ (right) scalings. The temperature color scale is identical for the middle and right panels.}\label{fig2}
\end{figure}

Fig.~\ref{fig1}, shows the local energy density (left) and local effective temperature (middle) as functions of the overlap order $k$. The right panel shows the average overlap order $\langle k(r) \rangle$ as a function of the radial distance from the origin for different $N_s$. Fig.~\ref{fig2} displays a transverse slice of a rare event with $N_s = 100$ strings: the number of overlaps (left) is converted into temperature maps using the $\varepsilon \propto \sqrt{k}$ (middle) and $\varepsilon \propto k$ (right) scalings, with a common color axis for direct comparison. When switching from the $T_k\propto k^{1/8}$ to the $T_k\propto k^{1/4}$ scaling, the highly overlapped regions become significantly hotter. However, both the radial profile and the event snapshot show that this dense core occupies only a small fraction of the string distribution.

The local temperature enters the energy-loss calculation both parametrically and through the medium gluon Debye mass. In our hybrid approach~\cite{Prokhorova:2025jmd}, the energy loss of a heavy-flavor (HF) quark passing through the string configuration is accumulated along its straight-line trajectory. Here, we estimate the average temperature an HF probe experiences along its path, while the full momentum-loss calculation will be presented elsewhere. This jet-path average $\langle T\rangle_\mathrm{jet}$ better captures the distinction between the dense core and the dilute halo than the area average $\langle T\rangle_\mathrm{area}$ (computed over all cells with $k\geq 1$).

\begin{figure}[h!]
\centering
\includegraphics[width=0.7\linewidth]{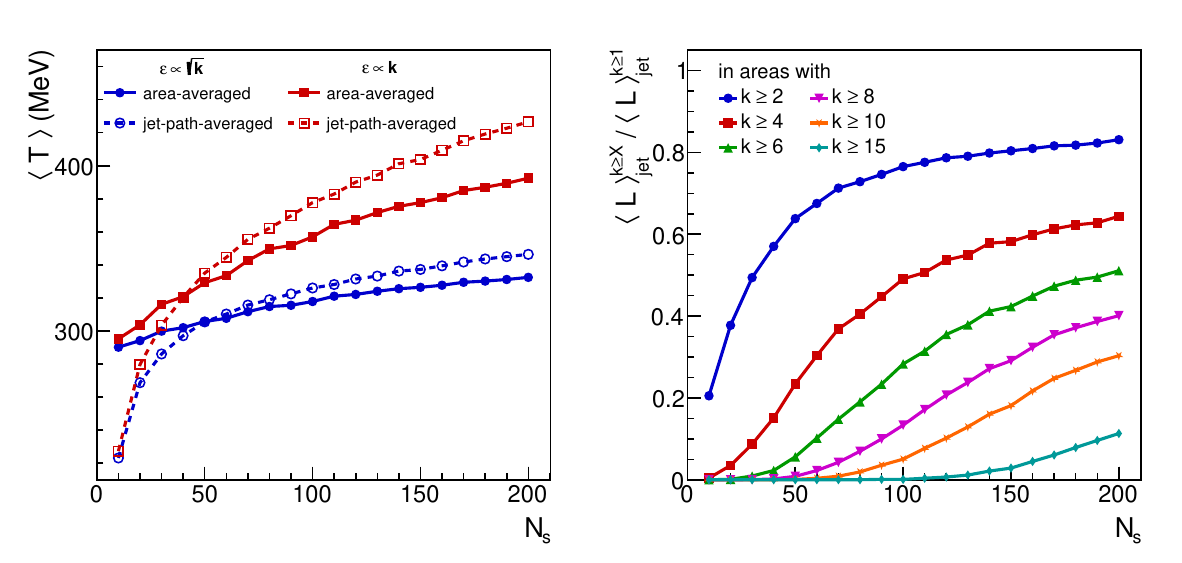}
\caption{Left: average effective temperature computed over the covered area (full symbols) and along jet trajectories (open symbols) for the $\sqrt{k}$ (blue) and $k$ (red) scalings. Right: fraction of the jet path spent in regions with $k \geq X$ relative to the total path in the medium. Both panels show the dependence on the event number of strings $N_s$.}\label{fig3}
\end{figure}

Jets are initialized with a probability proportional to the binary-collision density, which is biased toward the dense core. Directions are drawn uniformly and the probes are propagated along straight-lines. Along each trajectory $j$, we accumulate $I^{(j)} = \int dl \; T$ over cells with $k\geq1$ which also gives the in-medium path length $L^{(j)}$. The jet-path-averaged temperature is the mean of the individual path averages:
\begin{equation}
    \langle T\rangle_\mathrm{jet} = \frac{1}{N_\mathrm{jets}} \sum^{N_\mathrm{jets}}_{j=1} \frac{I^{(j)}}{L^{(j)}}.
\end{equation}
Fig.~\ref{fig3}, left, compares $\langle T\rangle_\mathrm{area}$ (full symbols) and $\langle T\rangle_\mathrm{jet}$ (open symbols) for both scalings. The right panel shows the fraction of the jet path spent in regions with $k \geq X$, i.e. $\langle L^{k\geq X}\rangle_\mathrm{jet}/\langle L^{k\geq 1}\rangle_\mathrm{jet}$, where all jet quantities are averaged over $10^4$ trajectories per $N_s$. At low $N_s$,  the string medium is dilute, and the area-averaged temperature exceeds the jet-path-averaged one. For $N_s \gtrsim 50$ this order reverses and the jet average shows a larger gap between the two scalings than the area average. This is explained by the path composition shown in the right panel: at $N_s = 100$, jets spend $\sim 80\%$ of their trajectory in $k \geq 2$ regions, and $\sim 20\%$ in $k \geq 8$ regions. Because the temperature difference between the scalings grows non-linearly with $k$ (Fig.~\ref{fig1}, middle), sampling these high-overlap regions amplifies the divergence along the probe's path, whereas the area average remains diluted by the vast $k=1$ halo even at high $N_s$ (Fig.~\ref{fig2}, left). 

\section{Conclusion}
\label{sec3}
We have compared two ways of extracting the medium energy density from an ensemble of overlapping color strings: identifying it with the field energy (giving $\varepsilon \propto k$) or assuming it inherits the field-strength scaling (giving $\varepsilon \propto \sqrt{k}$). Although these prescriptions yield very different dependencies of the local effective temperature on the overlap order (\(T_k\propto k^{1/4}\) and \(T_k\propto k^{1/8}\), respectively), the assumed localized nature of the high-overlap regions moderates their impact on spatially averaged observables. The presented comparison, therefore, probes both the spatial redistribution of energy and the overall energy available for the formation of the effective medium.

For our hybrid framework studying heavy-flavor probe energy loss, the jet-path-averaged effective temperature is a more relevant quantity than the area average, as it captures the difference between the dense core and the dilute halo along the probe's trajectory. Furthermore, observables sensitive to local fluctuations and spatial inhomogeneities, such as event-by-event HF energy loss, transverse-momentum broadening, and momentum imbalance or angular correlations of HF pairs, could provide even more sensitive probes of regions with varying string overlap.

\bibliographystyle{elsarticle-num}
\bibliography{xqcd2026_template}

\end{document}